**Quasi-Spin Ising Model and Monte Carlo Simulation of Ferroelastic Phase Transition: 3D Diffuse Scattering and Displacement Short-Range Ordering in Pre-Martensitic State**

Xiaoxu Guo,[1] Yongmei M. Jin,[1] Yang Ren,[2] and Yu U. Wang[1],*

[1] Department of Materials Science and Engineering, Michigan Technological University, Houghton, Michigan 49931, USA
[2] X-Ray Science Division, Advanced Photon Source, Argonne National Laboratory, Argonne, IL 60439, USA

**Abstract**

A quasi-spin Ising model of ferroelastic phase transition is developed and employed to perform atomic-scale Monte Carlo simulation of thermoelastic martensitic transformation. The quasi-spin variable associated with the lattice sites characterizes the local unit cells of the orientation variants of the ground-state martensite phase, which interact with each other through long-range elastic interactions. The simulation study focuses on the intrinsic behaviors of a defect-free crystal that undergoes cubic-to-tetragonal martensitic transformation. It is shown that the diffuse scattering in the pre-martensitic austenite state results from the spatial correlation of the atomic-scale heterogeneous lattice displacements and manifests the displacement short-range ordering. The effects of temperature, elastic anisotropy, and shear modulus softening on the diffuse scattering and displacement short-range ordering are investigated. It is found that elastic softening in the shear modulus $C' = (C_{11} - C_{12})/2$ promotes $\langle 110 \rangle |\langle 1\bar{1}0 \rangle$ displacement plane waves that stabilize the cubic austenite phase through increased entropy, decreasing the martensitic transformation temperature. The simulated diffuse scattering is compared and agrees with the complementary synchrotron X-ray single-crystal diffuse scattering experiment.

1. Introduction

A martensitic transformation is a first-order solid-state displacive (diffusionless) ferroelastic phase transition that breaks the crystal symmetry of high-temperature austenite phase by inducing spontaneous anisotropic lattice strain upon cooling and producing multiple low-symmetry orientation variants of low-temperature martensite phase [1-4]. Martensitic transformations are widely observed in a large group of materials including metal alloys and ceramics [5]. They have been extensively studied [1,2] for their fundamental importance in solid-state materials science and physics. They also provide the physical basis of shape memory effect, making some martensitic alloys important functional materials [6]. Despite successful applications of martensitic alloys as structural and functional materials, our fundamental understanding of martensitic transformations is still incomplete, and some important issues remain to be clarified [7-10]. One intriguing issue is pre-martensitic phenomena, also called martensite precursor effects. Prior to martensitic transformation, the high-symmetry cubic austenite phase usually undergoes incomplete phonon softening in a wide temperature range 10-100K above the martensite start temperature, which is accompanied by various anomalies that are unexpected in cubic phase [4,5,10,11]. In particular, strong diffuse scattering is observed in diffraction, which depends on the temperature and exhibits different characteristics around different Bragg reflection peaks [11]. Figure 1 shows the experimentally measured three-dimensional (3D) diffuse scattering around (800) Bragg reflection peak in $Ni_{49.90}Mn_{28.75}Ga_{21.35}$ single crystal at 490 K and 327 K, respectively, above the martensite start temperature 323 K. The measured 3D diffuse scattering around different Bragg reflection peaks $(HKL)$ at 330 K is shown in Figure 2. Details of the 3D diffuse scattering experiment using in-situ high-energy synchrotron X-ray single-crystal diffraction are described in Section 2.4.

---
* Author to whom correspondence should be addressed; electronic mail: wangyu@mtu.edu

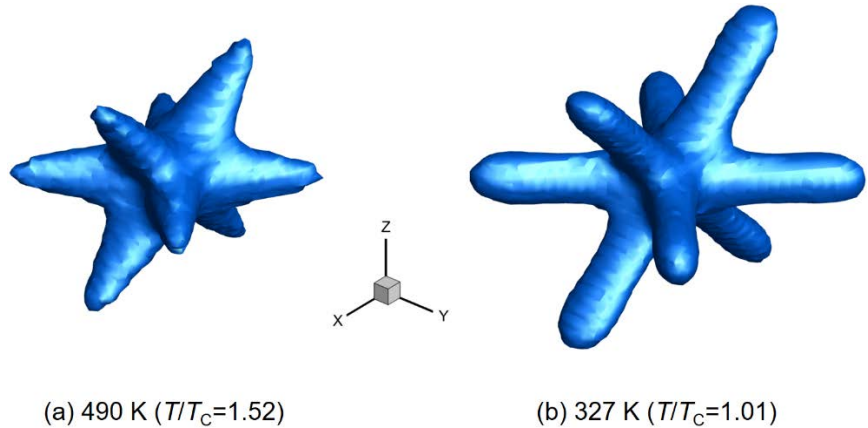

(a) 490 K ($T/T_C$=1.52)　　　　(b) 327 K ($T/T_C$=1.01)

**Figure 1.** Experimental measurement of 3D diffuse scattering around (800) Bragg reflection peak in $Ni_{49.90}Mn_{28.75}Ga_{21.35}$ at (a) 490 K and (b) 327 K using in-situ high-energy synchrotron X-ray single-crystal diffraction. The ferroelastic transition temperature (martensite start temperature) is $T_C$=323 K. The $x$, $y$ and $z$ axes are aligned along [100], [010] and [001] crystal axes, respectively.

　　The experimental phenomena exemplified in Figures 1 and 2 are intriguing because diffuse scattering manifests structural deviations from the cubic austenite crystal structure, and such deviations occur long before the first-order martensitic transformation (i.e., at temperature far above the martensite start temperature). Moreover, according to reciprocal relation, the intensity distribution of the diffuse scattering in the reciprocal lattice cells shown in Figure 2 implies an atomic lengthscale of the structural deviations with characteristic wavelengths of a few lattice parameters. Furthermore, repeated experiments reproduce the same diffuse scattering results at corresponding temperatures irrespective of prior thermal and mechanical histories, e.g., forward and reverse martensitic transformations, stress-induced martensitic transformation, annealing at temperatures up to 500 K, etc. These observations indicate that the diffuse scattering and underlying structural deviations are intrinsic properties of the cubic austenite phase in a thermodynamic equilibrium pre-martensitic state. In this paper, we focus on the intrinsic behaviors of a defect-free crystal that undergoes cubic↔tetragonal martensitic transformation. In particular, we carry out material modeling and computer simulation to correlate the diffuse scattering to the short-range ordering of atomic-scale heterogeneous displacement field developed in the pre-martensitic austenite crystal lattice prior to the development of long-range order during martensitic transformation.

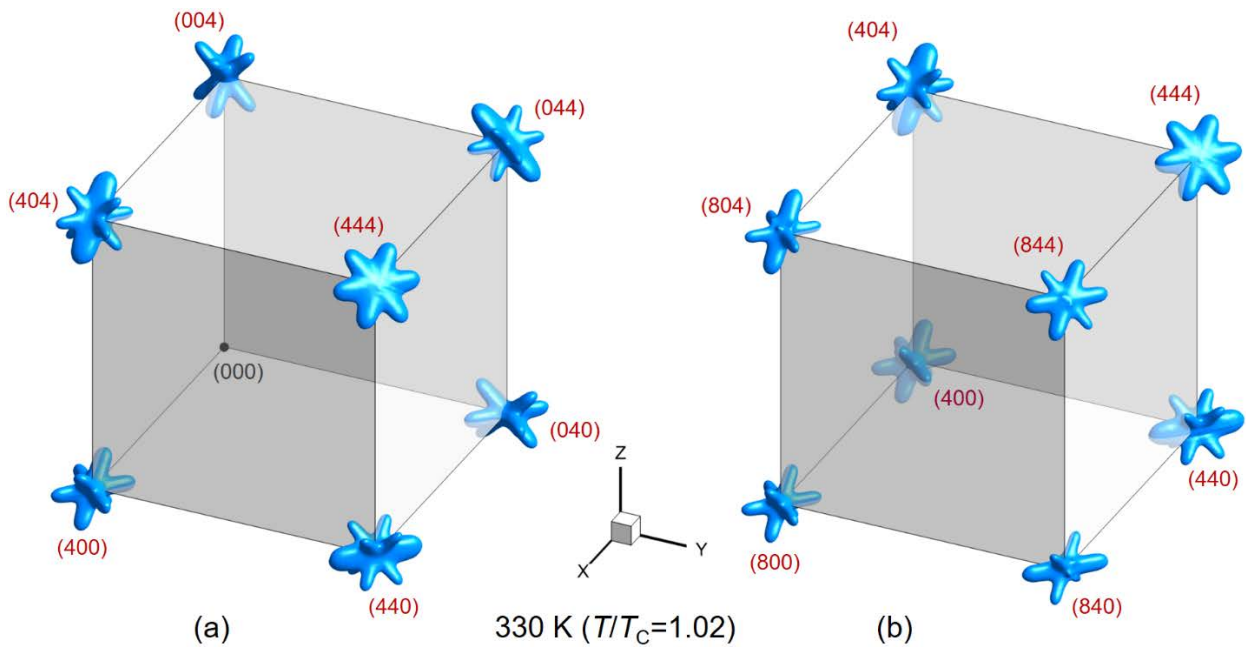

**Figure 2.** Experimental measurement of 3D diffuse scattering around different Bragg reflection peaks (*HKL*) in $Ni_{49.90}Mn_{28.75}Ga_{21.35}$ at 330 K using in-situ high-energy synchrotron X-ray single-crystal diffraction.

The above-discussed phenomena of diffuse scattering and atomic-scale heterogeneous lattice displacements dictate an atomic-scale modeling and simulation approach to martensitic transformations. While phase field microelasticity model is a well-established computational method to simulate microstructure evolution in martensitic transformations [12,13], its mesoscale nature makes it unsuitable for this task. Various atomic-scale computational methods have been employed to simulate martensitic transformations. Density Functional Theory has been employed to calculate the energetics of martensitic transformation along Bain path [14], but its limited supercell size and zero Kelvin computation make it unfit for direct simulation study of heterogeneous displacement field developed at finite temperatures in the pre-martensitic austenite crystal lattice. Molecular Dynamics has also been employed to simulate martensitic transformation [15], but its limited timescale (nanoseconds or shorter) and demanding computing requirements make it undesirable for comparing simulation results with the diffuse scattering experiments that measure the time-averaged intensities (over seconds or longer) of the weak diffuse scattering (the effects of time averaging of the simulated diffuse scattering intensity are discussed in Section 3.3). Atomic-scale Monte Carlo method covers sufficient length and time scales with affordable computing requirements and thus is ideal for simulation study of diffuse scattering and displacement short-range ordering at varying temperatures in the pre-martensitic austenite state. In this work, a quasi-spin Ising model of ferroelastic phase transition is developed and employed to perform atomic-scale Monte Carlo simulation of martensitic transformation.

It is worth noting that Monte Carlo simulation has been implemented in different manners based on different models. In one type of Monte Carlo simulations, Landau-type free energy density function is defined in terms of local structural order parameters (spontaneous strain components in the case of martensitic transformations), local disorders are introduced through coupling to the free energy expansion coefficient leading to deviations in local transformation temperature, and Monte Carlo simulation is used to minimize the total system free energy [16]. This type of model and Monte Carlo simulation is not adopted in our atomic-scale computational study of heterogeneous lattice displacements and diffuse scattering, because definitions of both free energy and transformation temperature are not well justified on a local unit cell level, and Monte Carlo simulation (i.e., Metropolis algorithm) based on the free energy change (rather than energy change) is inconsistent with statistical mechanics. To perform atomic-scale Monte Carlo simulation in this study, we adopt the Ising-type models with quasi-spin variables and generalized interactions specific to relevant physical problems. An Ising model with a fictitious spin variable associated with discrete lattice sites and nearest-neighbor effective $He^{\alpha}$-$He^{\beta}$ interactions has been developed to simulate the superfluid ordering and phase separation in $He^3$-$He^4$ mixtures [17]. An axial Ising model with a quasi-spin variable arranged in 1D array characterizing the stacking sequences of close-packed atomic layers and interlayer interaction potentials beyond nearest atomic layers has been employed to simulate stress-induced martensitic transformations in Cu-Al-Ni alloys [18]. A similar 1D Ising model has been used to simulate the evolution of diffuse scattering associated with the deformation-type stacking faults resulting from shear of close-packed planes during HCP to FCC martensitic transformation [19]. Besides the Monte Carlo simulations based on the Ising-type models with quasi-spin variables defined at lattice sites, off-lattice Monte Carlo method has also been employed to simulate martensitic transformation, where a randomly selected atom is displaced by a small random distance in a random direction during each Monte Carlo sampling trial and the energy change is evaluated from embedded atom method-derived interatomic potential [20]. Such an off-lattice Monte Carlo method extends the timescale of Molecular Dynamics at the expense of smoothing out thermal vibration dynamics, thus bridging the timescales of Molecular Dynamics and Ising lattice-based Monte Carlo methods. In this study, we adopt quasi-spin Ising model to carry out lattice-based Monte Carlo simulation, where the quasi-spin variable associated with the lattice sites characterizes the local unit cells of the orientation variants of the ground-state martensite phase, which interact with each other throughout all lattice sites as characterized by the long-range elastic interactions. This method is effective to treat the heterogeneous lattice displacements deviating from the average cubic lattice sites of the pre-martensitic austenite phase. As discussed in Sections 3.3 and 3.4, the heterogeneous lattice displacements treated in this manner are equivalent to displacement plane waves that mimic acoustic phonon branches relevant to the martensitic transformation. The computational and experimental methodologies are described in Section 2, and the simulation results are presented and discussed in Section 3.

## 2. Methodology

To gain insights into the intrinsic atomic-scale behaviors of a defect-free crystal that undergoes martensitic transformation, we develop a quasi-spin Ising model of ferroelastic phase transition (Section 2.1) in analogy to the Ising model of ferromagnetic phase transition. The model is employed to perform atomic-scale Monte Carlo simulation of thermoelastic martensitic transformation (Section 2.2). 3D diffuse scattering in the pre-martensitic austenite state is computed at different temperatures (Section 2.3) and compared with the complementary synchrotron X-ray single-crystal diffraction experiments (Section 2.4).

### 2.1. Quasi-Spin Ising Model of Ferroelastic Phase Transition and Thermoelastic Martensitic Transformation

In analogy to the spin variable in Ising model of ferromagnetism, the proposed Ising model of ferroelastic phase transition and thermoelastic martensitic transformation employs a multivalued quasi-spin variable $s(\mathbf{r})$ defined at each lattice site $\mathbf{r}$ to describe the system state. To be specific while without loss of generality, a martensitic system of tetragonal ground state is considered in the following. In such an exemplary system, the quasi-spin variable $s(\mathbf{r})$ assumes one of three values at each lattice site $\mathbf{r}$, namely, $s=1$, 2 or 3 characterizing the unit cell of tetragonal orientation variant 1, 2 or 3 of the ground-state martensite phase, respectively, each of which is associated with a stress-free lattice misfit strain $\varepsilon_{ij}^0(s)$:

$$\varepsilon_{ij}^0(1) = \begin{pmatrix} \varepsilon_c & 0 & 0 \\ 0 & \varepsilon_a & 0 \\ 0 & 0 & \varepsilon_a \end{pmatrix},\ \varepsilon_{ij}^0(2) = \begin{pmatrix} \varepsilon_a & 0 & 0 \\ 0 & \varepsilon_c & 0 \\ 0 & 0 & \varepsilon_a \end{pmatrix},\ \varepsilon_{ij}^0(3) = \begin{pmatrix} \varepsilon_a & 0 & 0 \\ 0 & \varepsilon_a & 0 \\ 0 & 0 & \varepsilon_c \end{pmatrix} \qquad (1)$$

For a given quasi-spin configuration described by the field variable $s(\mathbf{r})$, the lattice misfit strain field is:

$$\varepsilon_{ij}^0(\mathbf{r}) = \frac{1}{2}\varepsilon_{ij}^0(1)\left[2-s(\mathbf{r})\right]\left[3-s(\mathbf{r})\right] + \varepsilon_{ij}^0(2)\left[s(\mathbf{r})-1\right]\left[3-s(\mathbf{r})\right]$$
$$+ \frac{1}{2}\varepsilon_{ij}^0(3)\left[s(\mathbf{r})-1\right]\left[s(\mathbf{r})-2\right] \tag{2}$$

The elastic energy generated by this lattice misfit strain field $\varepsilon_{ij}^0(\mathbf{r})$ is [3]:

$$E_{el} = \frac{1}{2}\oint \frac{d^3k}{(2\pi)^3} B_{ijkl}(\mathbf{n})\tilde{\varepsilon}_{ij}^0(\mathbf{k})\tilde{\varepsilon}_{kl}^0(\mathbf{k})^* \tag{3}$$

where $B_{ijkl}(\mathbf{n}) = C_{ijkl} - n_p C_{pqij} \Omega_{qr}(\mathbf{n}) C_{rskl} n_s$, $C_{ijkl}$ is elastic modulus tensor, $\mathbf{k}$ is a vector in reciprocal space, $\mathbf{n} = \mathbf{k}/k$ is a unit directional vector in $\mathbf{k}$-space, $\Omega_{ij}(\mathbf{n})$ is the elastic Green function tensor inverse to $\Omega_{ij}^{-1}(\mathbf{n}) = C_{ikjl} n_k n_l$, $\tilde{\varepsilon}_{ij}^0(\mathbf{k})$ is the Fourier transform of the lattice misfit strain field $\varepsilon_{ij}^0(\mathbf{r})$, the superscript asterisk * indicates complex conjugate, summation convention over repeated indices is assumed, and $\oint$ is a principal-value integral in $\mathbf{k}$-space that is evaluated by excluding the point $\mathbf{k} = \mathbf{0}$.

While the above formulation of the Ising model of ferroelastic phase transition and thermoelastic martensitic transformation is straightforward, the lattice misfit strain field $\varepsilon_{ij}^0(\mathbf{r})$ in Eq. (2) is a nonlinear function of the quasi-spin variable $s(\mathbf{r})$, which prevents a simplification of the formulae for the elastic energy and strain field and thus hinders efficient algorithms of Monte Carlo simulation and diffuse scattering computation. Therefore, it is desirable to reformulate Eq. (2) in an equivalent but more convenient fashion. To this end, a three-component field variable $\hat{\eta} = (\eta_1, \eta_2, \eta_3)$ is adopted (note that $\hat{\eta}$ is not a vector), where each component describes one orientation variant, i.e., $(\eta_1, \eta_2, \eta_3) = (1,0,0)$, $(0,1,0)$ and $(0,0,1)$ correspond to $s=1$, 2 and 3, respectively, and thus one and only one component assumes value 1 at each lattice site while the other two components assume value 0. The lattice misfit strain field is a linear function of $\eta_i(\mathbf{r})$:

$$\varepsilon_{ij}^0(\mathbf{r}) = \varepsilon_{ij}^0(1)\eta_1(\mathbf{r}) + \varepsilon_{ij}^0(2)\eta_2(\mathbf{r}) + \varepsilon_{ij}^0(3)\eta_3(\mathbf{r}) \tag{4}$$

Upon substitution of Eq. (1), Eq. (4) becomes:

$$\varepsilon_{ij}^0(\mathbf{r}) = \varepsilon_a \delta_{ij} + \gamma_0 \begin{bmatrix} \eta_1(\mathbf{r}) & 0 & 0 \\ 0 & \eta_2(\mathbf{r}) & 0 \\ 0 & 0 & \eta_3(\mathbf{r}) \end{bmatrix} \tag{5}$$

where $\delta_{ij}$ is Kronecker delta, $\gamma_0 = \varepsilon_c - \varepsilon_a$, and the identity $\eta_1 + \eta_2 + \eta_3 = 1$ at every lattice site $\mathbf{r}$ has been used. The first term in Eq. (5) represents an isotropic volume strain that is homogeneous throughout the whole system and thus does not generate lattice misfit — it does not contribute to the elastic energy in Eq. (3) since its Fourier transform is nonzero only at $\mathbf{k} = \mathbf{0}$ which is excluded in the principal-value integral.

Substituting Eq. (5) into Eq. (3) reduces the elastic energy formula, which can be expressed explicitly in terms of $\tilde{\eta}_i(\mathbf{k})$, the Fourier transform of $\eta_i(\mathbf{r})$, when elastic modulus tensor of cubic symmetry is considered. Instead of the three cubic elastic constants $C_{11}$, $C_{12}$ and $C_{44}$, it is more convenient to formulate in terms of shear modulus $C' = (C_{11} - C_{12})/2$ on (110) plane, Poisson's ratio $v = C_{12}/(C_{11} + C_{12})$ under [100] uniaxial stress, and elastic anisotropy parameter $A = 1 + a = C/C'$ defined as the ratio between shear modulus $C = C_{44}$ on (100) plane and shear modulus $C'$ on (110) plane. The elastic energy is expressed as:

$$E_{el} = C' \gamma_0^2 \oint \frac{d^3 k}{(2\pi)^3} \frac{F(\mathbf{k})}{D(\mathbf{n})} \tag{6}$$

where $D(\mathbf{n})$ is defined as:

$$D(\mathbf{n}) = (1-v)(1+a)^2 - 2a(1+a)(n_1^2 n_2^2 + n_2^2 n_3^2 + n_3^2 n_1^2) + 2a^2 [3 + (1-2v)a] n_1^2 n_2^2 n_3^2 \tag{7}$$

and $F(\mathbf{k})$ is an explicit function of $\tilde{\eta}_i(\mathbf{k})$:

$$F(\mathbf{k}) = \sum_{i=1}^{3} \sum_{j=1}^{3} \Phi_{ij}(\mathbf{n}) \tilde{\eta}_i(\mathbf{k}) \tilde{\eta}_j(\mathbf{k})^* \tag{8}$$

where $\Phi_{ij}(\mathbf{n})$ has the following components (note that $\Phi_{ij}$ is not a tensor):

$$\Phi_{11}(\mathbf{n}) = (1+a)\{(1+a)(1-n_1^2)^2 - 2a(1+v)n_2^2 n_3^2 + 2a[2+a(1-v)]n_1^2 n_2^2 n_3^2\}$$

$$\Phi_{22}(\mathbf{n}) = (1+a)\{(1+a)(1-n_2^2)^2 - 2a(1+v)n_3^2 n_1^2 + 2a[2+a(1-v)]n_1^2 n_2^2 n_3^2\}$$

$$\Phi_{33}(\mathbf{n}) = (1+a)\{(1+a)(1-n_3^2)^2 - 2a(1+v)n_1^2 n_2^2 + 2a[2+a(1-v)]n_1^2 n_2^2 n_3^2\} \tag{9}$$

$$\Phi_{23}(\mathbf{n}) = \Phi_{32}(\mathbf{n}) = (1+a)[(1+a)v n_1^2 + (1+a)n_2^2 n_3^2 - 2a(1-av)n_1^2 n_2^2 n_3^2]$$

$$\Phi_{31}(\mathbf{n}) = \Phi_{13}(\mathbf{n}) = (1+a)[(1+a)v n_2^2 + (1+a)n_3^2 n_1^2 - 2a(1-av)n_1^2 n_2^2 n_3^2]$$

$$\Phi_{12}(\mathbf{n}) = \Phi_{21}(\mathbf{n}) = (1+a)[(1+a)v n_3^2 + (1+a)n_1^2 n_2^2 - 2a(1-av)n_1^2 n_2^2 n_3^2]$$

In the case of isotropic elasticity, $a = 0$ (i.e., $A = 1$), Eq. (7) reduces to $D(\mathbf{n}) = 1 - v$, and Eq. (8) simplifies to:

$$\begin{aligned}
F(\mathbf{k}) = &\left(1-n_1^2\right)^2 \tilde{\eta}_1(\mathbf{k})\tilde{\eta}_1(\mathbf{k})^* + \left(1-n_2^2\right)^2 \tilde{\eta}_2(\mathbf{k})\tilde{\eta}_2(\mathbf{k})^* + \left(1-n_3^2\right)^2 \tilde{\eta}_3(\mathbf{k})\tilde{\eta}_3(\mathbf{k})^* \\
&+ \left\{vn_1^2 + n_2^2 n_3^2\right\}\left[\tilde{\eta}_2(\mathbf{k})\tilde{\eta}_3(\mathbf{k})^* + \tilde{\eta}_2(\mathbf{k})^* \tilde{\eta}_3(\mathbf{k})\right] \\
&+ \left\{vn_2^2 + n_3^2 n_1^2\right\}\left[\tilde{\eta}_3(\mathbf{k})\tilde{\eta}_1(\mathbf{k})^* + \tilde{\eta}_3(\mathbf{k})^* \tilde{\eta}_1(\mathbf{k})\right] \\
&+ \left\{vn_3^2 + n_1^2 n_2^2\right\}\left[\tilde{\eta}_1(\mathbf{k})\tilde{\eta}_2(\mathbf{k})^* + \tilde{\eta}_1(\mathbf{k})^* \tilde{\eta}_2(\mathbf{k})\right]
\end{aligned} \quad (10)$$

Here it is worth making a brief conceptual comparison between the above formulated model with the Ising model of ferromagnetism. The role of exchange interaction between nearest-neighbor spin pairs in the Ising model of ferromagnetism is replaced by the long-range dipole-dipole-like elastic interactions among the orientation variants over all lattice sites in the above quasi-spin Ising model of ferroelastic phase transition and thermoelastic martensitic transformation. At sufficiently low temperature, the ferromagnetic exchange interactions (with positive sign) bring the system into an ordered spin configuration of aligned spins, i.e., ferromagnetic phase; similarly, the elastic interactions (which always produce a nonnegative total elastic energy) bring the system into an ordered quasi-spin configuration described by the field variable $s(\mathbf{r})$, which characterizes martensitic phase. At sufficiently high temperature, thermal fluctuations disturb the spins into disordered configurations and each site possesses a zero average magnetic dipole moment, i.e., paramagnetic phase; similarly, disordered quasi-spin configurations described by the time-dependent field variable $s(\mathbf{r};t)$ are expected to result from thermal fluctuations, where each lattice site frequently switches among the three orientation variants and thus the unit cell possesses an average cubic symmetry, i.e., austenitic phase. In addition to the phase transitions occurring in both models, there is another aspect of fundamental importance in the analogy: a strong correlation develops among the spins at temperature close to but above the Curie point of ferromagnetic transition, which reflects the short-range ordering in the paramagnetic state; similarly, displacement short-range ordering is expected in the austenitic phase, which could shed light on the pre-martensitic phenomena. In particular, this proposed quasi-spin Ising model of ferroelastic phase transition and thermoelastic martensitic transformation will be employed to investigate the displacement short-range ordering phenomenon. To this end, diffuse scattering produced by the short-range ordered displacement field will be computed for the simulated quasi-spin configuration $s(\mathbf{r};t)$, as described in Section 2.3.

## 2.2. Monte Carlo Simulation

The elastic energy of the quasi-spin configuration $s(\mathbf{r})$ or equivalently $\hat{\eta}(\mathbf{r}) = [\eta_1(\mathbf{r}), \eta_2(\mathbf{r}), \eta_3(\mathbf{r})]$ in Eqs. (3) and (6) is the Hamiltonian function, which is used to evolve the quasi-spin configuration and explore equilibrium state at given temperature via Monte

Carlo simulation using Metropolis algorithm [21]. In each sampling trial, one lattice site is randomly selected, its current orientation variant is switched randomly into one of the other two variants, and the acceptance ratio of this attempt is given by the probability ratio of the Boltzmann factors of the attempted and current configurations: $\alpha = \exp(-\Delta E_{el}/k_B T)$, where $k_B$ is Boltzmann constant, $T$ is absolute temperature, and $\Delta E_{el}$ is the elastic energy change associated with this attempted variant switching on the selected lattice site.

According to Eq. (8), in order to evaluate the elastic energy change $\Delta E_{el}$ associated with the orientation variant switching on a lattice site $\mathbf{r}'$, the Fourier transforms $\tilde{\eta}_i(\mathbf{k})$ and $\tilde{\eta}'_i(\mathbf{k})$ of the affected field variables $\eta_i(\mathbf{r})$ before and $\eta'_i(\mathbf{r})$ after the attempted switching are needed to calculate the respective energies. To be specific, let us consider $\eta_I(\mathbf{r}')$ which switches from 1 to 0, and $\eta_J(\mathbf{r}')$ which switches from 0 to 1, while the third field is not affected. Before the attempted switching, their Fourier transforms are $\tilde{\eta}_I(\mathbf{k})$ and $\tilde{\eta}_J(\mathbf{k})$, respectively; after the above considered variant switching at lattice site $\mathbf{r}'$, the Fourier transforms $\tilde{\eta}'_I(\mathbf{k})$ and $\tilde{\eta}'_J(\mathbf{k})$ can be evaluated from $\tilde{\eta}_I(\mathbf{k})$ and $\tilde{\eta}_J(\mathbf{k})$ using the following updating formulae:

$$\tilde{\eta}'_I(\mathbf{k}) = \tilde{\eta}_I(\mathbf{k}) - e^{-i\mathbf{k}\cdot\mathbf{r}'}, \quad \tilde{\eta}'_J(\mathbf{k}) = \tilde{\eta}_J(\mathbf{k}) + e^{-i\mathbf{k}\cdot\mathbf{r}'} \tag{11}$$

which is more efficient than performing discrete Fourier transform for $\eta'_I(\mathbf{r})$ and $\eta'_J(\mathbf{r})$. Therefore, only the initial fields $\tilde{\eta}_i(\mathbf{k};t_0)$ need to be evaluated by fast Fourier transform, while the subsequent fields can be recursively updated by using Eq. (11) during the Monte Carlo simulation.

In terms of discrete Fourier transform, the $\mathbf{k}$-space integration in Eq. (6) becomes summation over all $\mathbf{k}$ points within the first Brillouin zone of the reciprocal lattice associated with the lattice sites $\mathbf{r}$ in real space:

$$E_{el} = C'\gamma_0^2 \frac{V}{N^2} \sum_{\mathbf{k}}' \frac{F(\mathbf{k})}{D(\mathbf{n})} \tag{12}$$

where $V$ is the total system volume, $N = N_x N_y N_z$ is the total number of lattice sites, and $\sum_{\mathbf{k}}'$ excludes $\mathbf{k} = \mathbf{0}$. According to Eqs. (7-9) and (12), the elastic energy $E_{el}$ before the attempted switching is evaluated using $\tilde{\eta}_I(\mathbf{k})$ and $\tilde{\eta}_J(\mathbf{k})$ (together with the third unaffected field), the elastic energy $E'_{el}$ after the attempted switching is evaluated using $\tilde{\eta}'_I(\mathbf{k})$ and $\tilde{\eta}'_J(\mathbf{k})$, and the energy change associated with this attempted switching is $\Delta E_{el} = E'_{el} - E_{el}$, which is used to determine the acceptance ratio of the attempt in each Monte Carlo sampling trial.

## 2.3. Computational Diffuse Scattering

The lattice misfit strain field $\varepsilon_{ij}^0(\mathbf{r})$ given in Eq. (5) generates a displacement field $u_i(\mathbf{r})$ distributed on the lattice sites $\mathbf{r}$, which can be expressed as a sum of a linear function of $\mathbf{r}$ and a heterogeneous displacement field $v_i(\mathbf{r})$, i.e., $u_i(\mathbf{r}) = \overline{\varepsilon}_{ij} r_j + v_i(\mathbf{r})$. The symmetric tensor $\overline{\varepsilon}_{ij} = \overline{\varepsilon}_{ji}$ describes a homogeneous strain throughout the system and is given by:

$$\overline{\varepsilon}_{ij} = \overline{\varepsilon}_{ij}^0 = \varepsilon_a \delta_{ij} + \gamma_0 \left( \overline{\eta}_1 \delta_{i1} \delta_{j1} + \overline{\eta}_2 \delta_{i2} \delta_{j2} + \overline{\eta}_3 \delta_{i3} \delta_{j3} \right) \tag{13}$$

where $\overline{\eta}_i = V^{-1} \int_V \eta_i(\mathbf{r}) d^3 r = N^{-1} \sum_{\mathbf{r}} \eta_i(\mathbf{r})$ is the spatial average of $\eta_i(\mathbf{r})$ over the system volume. The homogeneous strain $\overline{\varepsilon}_{ij}$ determines the positions of Bragg reflection peaks from the quasi-spin configuration $s(\mathbf{r})$ or equivalently $\hat{\eta}(\mathbf{r}) = [\eta_1(\mathbf{r}), \eta_2(\mathbf{r}), \eta_3(\mathbf{r})]$. As shown in Section 3.2, the temperature-dependent thermal fluctuations during ferroelastic phase transition and thermoelastic martensitic transformation can be monitored by the macroscopic average strain $\overline{\varepsilon}_{ij}$, i.e., the three principal components $\overline{\varepsilon}_{11} = \varepsilon_a + \gamma_0 \overline{\eta}_1$, $\overline{\varepsilon}_{22} = \varepsilon_a + \gamma_0 \overline{\eta}_2$ and $\overline{\varepsilon}_{33} = \varepsilon_a + \gamma_0 \overline{\eta}_3$.

The heterogeneous displacement field $v_i(\mathbf{r})$ produces local heterogeneous strain in the lattice, which describes the structural deviations from the homogeneous average lattice structure. The Fourier transform $\tilde{v}_i(\mathbf{k})$ of the heterogeneous displacement field $v_i(\mathbf{r})$ is given by [3]:

$$\tilde{v}_i(\mathbf{k}) = -\frac{i}{k} \Omega_{ij}(\mathbf{n}) C_{jklm} n_k \Delta \tilde{\varepsilon}_{lm}^0(\mathbf{k}) \tag{14}$$

where $\Delta \varepsilon_{ij}^0(\mathbf{r}) = \varepsilon_{ij}^0(\mathbf{r}) - \overline{\varepsilon}_{ij}^0$ and thus $\Delta \tilde{\varepsilon}_{ij}^0(\mathbf{k} = \mathbf{0}) = 0$ and $\tilde{v}_i(\mathbf{k} = \mathbf{0}) = 0$. Using Eq. (5) and elastic modulus tensor of cubic symmetry, Eq. (14) reduces to:

$$\tilde{v}_i(\mathbf{k}) = \gamma_0 \frac{i}{k} \frac{\Psi_i(\mathbf{k})}{D(\mathbf{n})} n_i \tag{15}$$

where no summation is assumed over the repeated index $i$, $D(\mathbf{n})$ is the same as defined in Eq. (7), and $\Psi_i(\mathbf{k})$ is (note that $\Psi_i$ is not a vector):

$$\begin{aligned}\Psi_1(\mathbf{k}) &= \left[ (1-v)(1+a)^2 + (1-a^2)(n_2^2 + n_3^2) - 4a(1-va) n_2^2 n_3^2 \right] \tilde{\eta}_1(\mathbf{k}) \\ &+ (1+a) \left[ v(1+a) - (1+a) n_2^2 + 2a n_3^2 (n_2^2 - v) \right] \tilde{\eta}_2(\mathbf{k}) \\ &+ (1+a) \left[ v(1+a) - (1+a) n_3^2 + 2a n_2^2 (n_3^2 - v) \right] \tilde{\eta}_3(\mathbf{k}) \end{aligned} \tag{16a}$$

$$\Psi_2(\mathbf{k}) = (1+a)\left[v(1+a) - (1+a)n_1^2 + 2an_3^2(n_1^2 - v)\right]\tilde{\eta}_1(\mathbf{k})$$
$$+ \left[(1-v)(1+a)^2 + (1-a^2)(n_3^2 + n_1^2) - 4a(1-va)n_3^2n_1^2\right]\tilde{\eta}_2(\mathbf{k}) \qquad (16b)$$
$$+ (1+a)\left[v(1+a) - (1+a)n_3^2 + 2an_1^2(n_3^2 - v)\right]\tilde{\eta}_3(\mathbf{k})$$

$$\Psi_3(\mathbf{k}) = (1+a)\left[v(1+a) - (1+a)n_1^2 + 2an_2^2(n_1^2 - v)\right]\tilde{\eta}_1(\mathbf{k})$$
$$+ (1+a)\left[v(1+a) - (1+a)n_2^2 + 2an_1^2(n_2^2 - v)\right]\tilde{\eta}_2(\mathbf{k}) \qquad (16c)$$
$$+ \left[(1-v)(1+a)^2 + (1-a^2)(n_1^2 + n_2^2) - 4a(1-va)n_1^2n_2^2\right]\tilde{\eta}_3(\mathbf{k})$$

which can be further simplified in the case of elastic isotropy, i.e., $a = 0$.

An existence of the heterogeneous displacement field $v_i(\mathbf{r})$ in the lattice produces diffuse scattering, where the scattering intensity is distributed in the reciprocal space at positions away from the Bragg reflection peaks. In particular, the diffuse scattering intensity distribution around Bragg peak centered at fundamental reciprocal lattice vector $\mathbf{K}$ is given by [22]:

$$I_{\text{diff}}(\mathbf{K};\mathbf{k}) = V_{\text{cell}}^{-2}|\tilde{n}_0|^2|(\mathbf{K}+\mathbf{k})\cdot\tilde{\mathbf{v}}(\mathbf{k})|^2 \propto |(\mathbf{K}+\mathbf{k})\cdot\tilde{\mathbf{v}}(\mathbf{k})|^2 \qquad (17)$$

where $V_{\text{cell}}$ is the unit cell volume of the lattice, $\tilde{n}_0$ is the structure factor, the reciprocal lattice vector $\mathbf{K}$ can be expressed via the Bragg peak index $(HKL)$, and $\mathbf{k}$ is defined within the first Brillouin zone centered at $\mathbf{K}$. For a cubic array of lattice sites $\mathbf{r}$, $\mathbf{K} \propto (H, K, L)$. The diffuse scattering exhibits different intensity distribution features around different Bragg peaks through the dependence on the peak indices $(HKL)$, as shown in Section 3.4.

## 2.4. 3D Diffuse Scattering Using In-Situ High-Energy Synchrotron X-Ray Single-Crystal Diffraction

To complement the simulation study, 3D diffuse scattering from thermoelastic martensitic Ni-Mn-Ga alloy single crystals is measured experimentally at temperatures above the martensitic transformation by using in-situ high-energy synchrotron X-ray diffraction. The diffuse scattering experiments are performed at the beamline 11-ID-C of the Advanced Photon Source at Argonne National Laboratory. High-flux high-energy synchrotron X-ray single-crystal diffraction is essential for such experiments: the high energy (115 keV) provides deep penetration to probe bulk single crystal specimens, the corresponding short wavelength (0.107805 Å) enables coverage of large volume in reciprocal space to measure high-index peaks, the high flux enables short exposure time (seconds for each exposure) to measure weak diffuse scattering intensity that is critical for 3D reciprocal space mapping, and 3D single-crystal diffraction allows separation of different diffuse scattering rods associated with displacement correlations in different directions — 2D measurement using rocking crystal method causes intensity overlapping problem where diffuse scattering intensities in different directions (such as along 12 <110> axes) are all projected into one image. A beamsize of 400×400 μm$^2$ cross-section is used. Scattering intensity data are measured by 2D digital X-ray detector (PerkinElmer® XRD 1621 AN, 2048×2048 pixels, 200 μm pixel size, 16 bit digitization). Temperature is controlled in-situ by using cryostream (Oxford Cryosystems® 700 Plus). To scan the 3D reciprocal space, the single crystal specimen is rotated incrementally by $\Delta\omega$=0.1° during each frame exposure. At each rotation position $\omega$ (defined at the midpoint of each increment $\Delta\omega$), the 2D detector measures the scattering intensity distribution that intersects the Ewald sphere. The 3D reciprocal space map of diffuse scattering intensity is constructed from a series of such 2D data. Details of the experiments are reported elsewhere [11,23]. Figure 1 shows the measured 3D diffuse scattering around (800) Bragg reflection peak in Ni$_{49.90}$Mn$_{28.75}$Ga$_{21.35}$ at 490 K and 327 K, respectively, above the martensite start temperature 323 K. The measured 3D diffuse scattering around different Bragg reflection peaks $(HKL)$ at 330 K is shown in Figure 2.

## 3. Results and Discussions

To carry out Monte Carlo simulation based on the above formulated quasi-spin Ising model of ferroelastic phase transition and thermoelastic martensitic transformation, it is convenient to introduce normalized dimensionless temperature. To this end, an energy unit is defined as $E_0 = C\gamma_0^2 \Delta V$, where $C$ is an elastic constant, and $\Delta V = V/N$ is the unit cell volume of the lattice. A reference temperature is defined as $T_0 = E_0/k_B$, and the dimensionless temperature $T^*$ is defined by normalizing the temperature $T$ with respect to the reference temperature $T_0$, i.e., $T^* = T/T_0$. In the following, the simulation results are presented in $T^*$. It is worth noting that the shear modulus $C = C_{44}$ on (100) plane, rather than the shear modulus $C' = (C_{11} - C_{12})/2$ on (110) plane, is used in the normalization, because $C'$ usually exhibits significant temperature-dependent softening in pre-martensitic austenite while $C$ remains approximately constant. In the simulations presented in the following, 32×32×32 computational grids are employed, and isotropic elasticity is assumed, i.e., $a = 0$ ($A = 1$), except for Section 3.6 which addresses the effects of elastic anisotropy on the phase transition and diffuse scattering.

### 3.1. Phase Transition

As expected, a phase transition occurs in the above formulated quasi-spin Ising model. Figure 3 shows the simulated equilibrium states representative of disordered quasi-spin configuration at $T^* = 0.22$ and ordered configuration at $T^* = 0.20$, respectively. In the disordered state shown in Figure 3(a), the three martensite orientation variants appear to randomly distribute throughout the system, where the crystal lattice averaged over the system volume is cubic (i.e., $\bar{\eta}_i = \frac{1}{3}$), and the local unit cell averaged over the simulation time is also cubic (i.e., $\langle \eta_i(\mathbf{r}) \rangle = \frac{1}{3}$); nevertheless, a quantitative analysis reveals an existence of spatial correlation among these variants at temperatures above the phase transition, which reflects temperature-dependent displacement short-range ordering in the pre-martensitic austenite, as discussed in more detail in Section 3.3. In the ordered state shown in Figure 3(b), the orientation variants self-assemble into polytwinned plates characteristic of martensitic microstructures, which reflects development of long-range order in the martensite. To visualize the elastic distortions in the lattice due to coherency strain, Figure 3 also shows the deformed lattices defined by the displaced sites $\mathbf{r} + \mathbf{u}(\mathbf{r})$ at the corresponding temperatures, where the austenite maintains an average cubic lattice (averaged over volume or over simulation time), while the martensitic lattice exhibits characteristic surface relief features associated with the martensite plates (such a martensitic microstructure corresponds to a low-energy quasi-spin configuration that is stable at low temperature with respect to time).

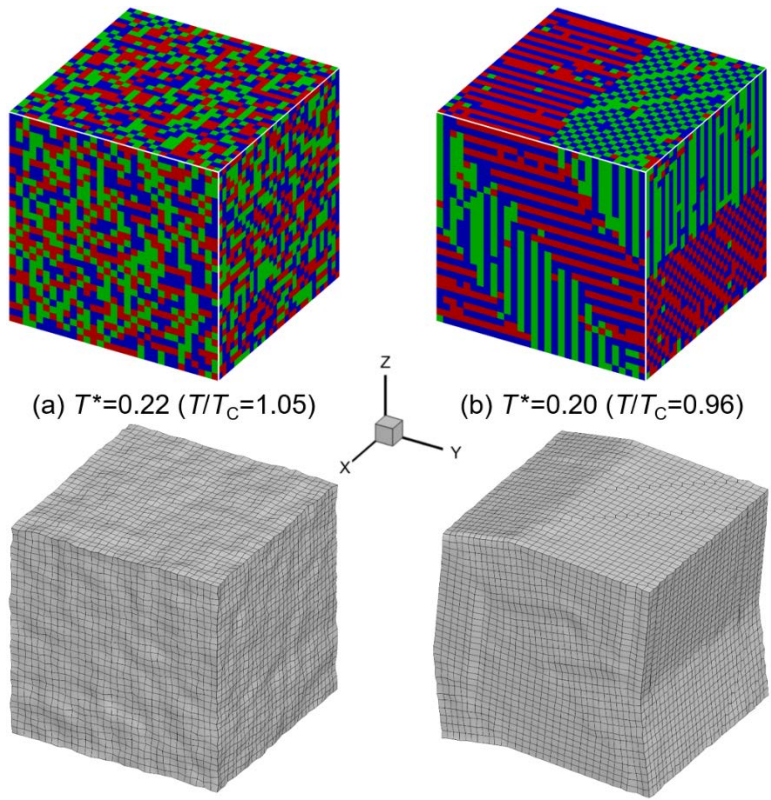

**Figure 3.** Simulated equilibrium multi-variant configurations representative of (a) disordered austenitic state at $T=1.05T_C$ and (b) ordered martensitic state at $T=0.96T_C$. Top row illustrates spatial distributions of three orientation variants by three colors (red, green, blue). Bottom row illustrates elastic distortion in the lattice due to coherency strain (for clarity, $\gamma_0 = \varepsilon_c - \varepsilon_a = 0.4$ is assumed).

The simulation results shown in Figure 3 indicate a phase transition temperature between $T^* = 0.22$ and $T^* = 0.20$. As shown in Section 3.2, the phase transition temperature is more precisely determined to be $T_C^* = 0.209$ through quantitative analysis of the temperature-dependent thermal fluctuations in strains near the phase transition. Figure 4 shows the simulated evolution process during the phase transition when an equilibrium disordered configuration obtained at $T^* = 0.22$ is quenched to and held at $T^* = 0.20$. At temperature above $T_C^*$, thermal fluctuations disturb the orientation variant distribution into disordered configurations, where each lattice site frequently switches among the three orientation variants and thus the average lattice symmetry is cubic, i.e., austenite phase. At temperature below $T_C^*$, thermal fluctuations are inadequate to compete with the long-range dipole-dipole-like elastic interactions among the orientation variants, which bring the system into an ordered configuration, i.e., martensite phase. The ordered orientation variant arrangement, i.e., polytwinned plate microstructure, self-accommodates the lattice misfit among the orientation variants and minimizes the elastic energy of the whole system. As shown in Figure 4(a-e), after being quenched to below $T_C^*$, the original disordered configuration becomes unstable, the orientation variants start to develop strong spatial correlation over longer distance, and eventually form long-range ordered configuration.

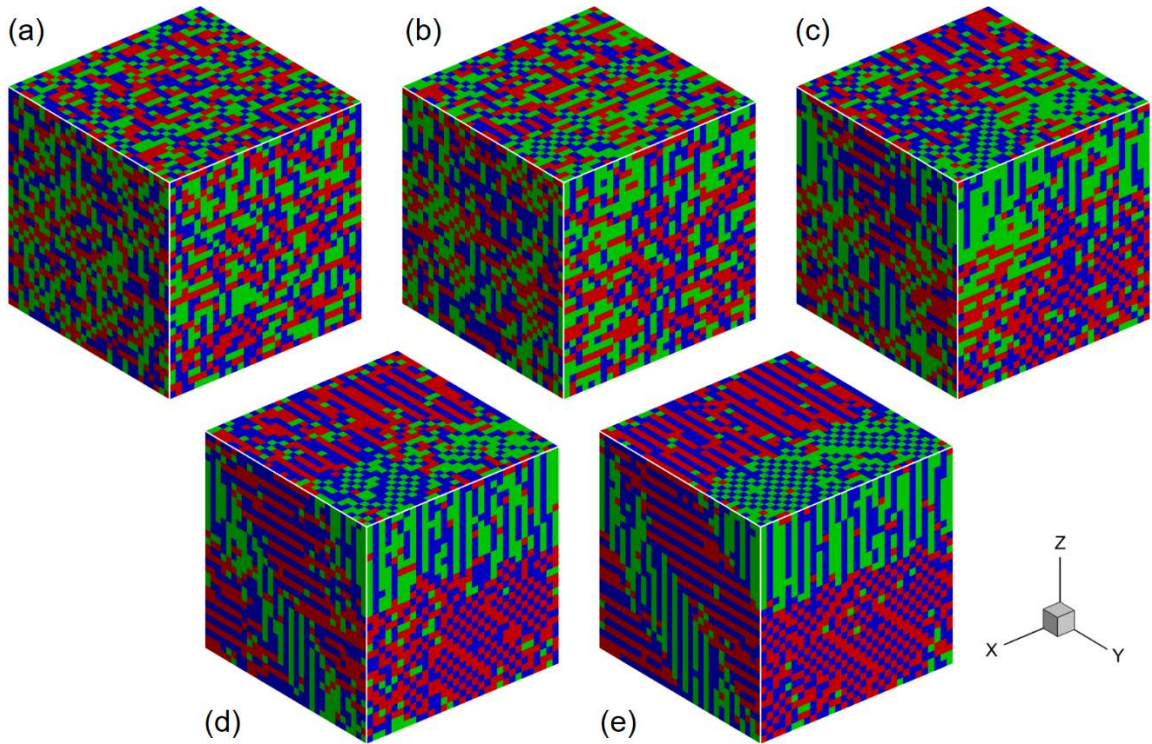

**Figure 4.** Simulated evolution process during martensitic phase transformation when an equilibrium disordered austenite configuration obtained at $T^* = 0.22$ ($T=1.05T_C$) is quenched to and held at $T^* = 0.20$ ($T=0.96T_C$) after (a) 0, (b) 50, (c) 100, (d) 150, and (e) 200 Monte Carlo steps.

## 3.2. Thermal Fluctuations near Phase Transition Temperature

According to fluctuation-dissipation theorem [24], the thermal fluctuation of the long-range order parameter (i.e., mean square fluctuation) of the phase transition is proportional to $(T - T_C)^{-1}$ in the vicinity of the phase transition temperature $T_C$. This relation allows a more precise determination of $T_C$ based on simulation results at a limited number of temperatures near $T_C$. In ferroelastic phase transition and thermoelastic martensitic transformation, which are displacive structural transformations, strain is the long-range order parameter. Thermal fluctuations in the macroscopic strain $\bar{\varepsilon}_{ij}^* = \bar{\varepsilon}_{ij}/\gamma_0$, in particular, the three principal components $\bar{\varepsilon}_{11}^*(t^*)$, $\bar{\varepsilon}_{22}^*(t^*)$ and $\bar{\varepsilon}_{33}^*(t^*)$ as described in Section 2.3, are simulated at temperatures above $T_C$, and two examples are plotted as function of the simulation time $t^*$ (i.e., Monte Carlo steps) in Figure 5(a,b). As expected, the strain fluctuations increase when $T$ decreases towards $T_C$. The standard deviation $\Delta\varepsilon_{ij}^*$ in each strain component $\bar{\varepsilon}_{ij}^*(t^*)$ is calculated with respect to its time-average value $\langle\bar{\varepsilon}_{ij}^*(t^*)\rangle$, and the values of $(\Delta\varepsilon^*)^2 = (\Delta\varepsilon_{11}^*)^2 + (\Delta\varepsilon_{22}^*)^2 + (\Delta\varepsilon_{33}^*)^2$ and its reciprocal $(\Delta\varepsilon^*)^{-2}$ are plotted as function of the temperature $T^*$ in Figure 5(c). A linear fitting of the data points in the vicinity of $T_C$ in the functional form of $(\Delta\varepsilon)^2 \propto (T - T_C)^{-1}$ yields $T_C^* = 0.209$, as shown in Figure 5(c), which is in agreement with the simulation results presented in Figure 3. In the discussions of the simulation results, it is more meaningful to consider the "homologous" temperature $T_H = T^*/T_C^* = T/T_C$, which is provided in the figures and/or captions presented in this paper.

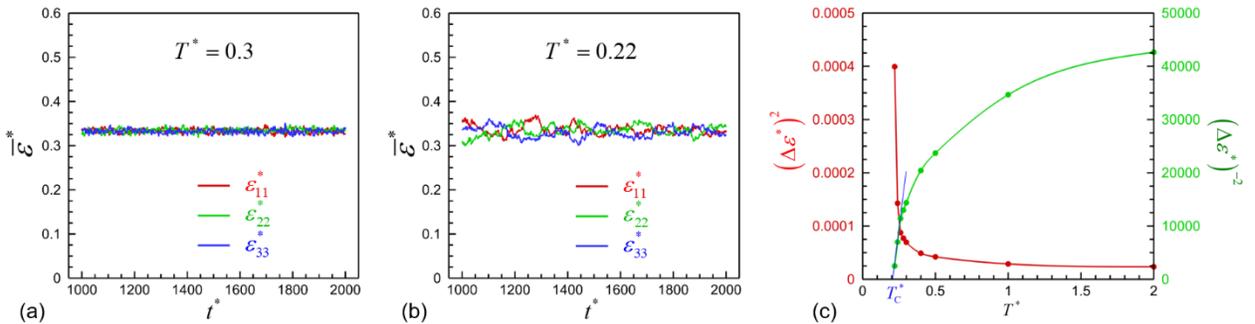

**Figure 5.** Simulated fluctuations of strain components at representative temperatures above phase transition temperature: (a) $T^* = 0.3$ and (b) $T^* = 0.22$. (c) Strain fluctuation as a function of temperature. The blue line represents linear fitting in the vicinity of the phase transition temperature, which gives $T_{\mathrm{C}}^* = 0.209$.

## 3.3. Diffuse Scattering and Displacement Short-Range Ordering

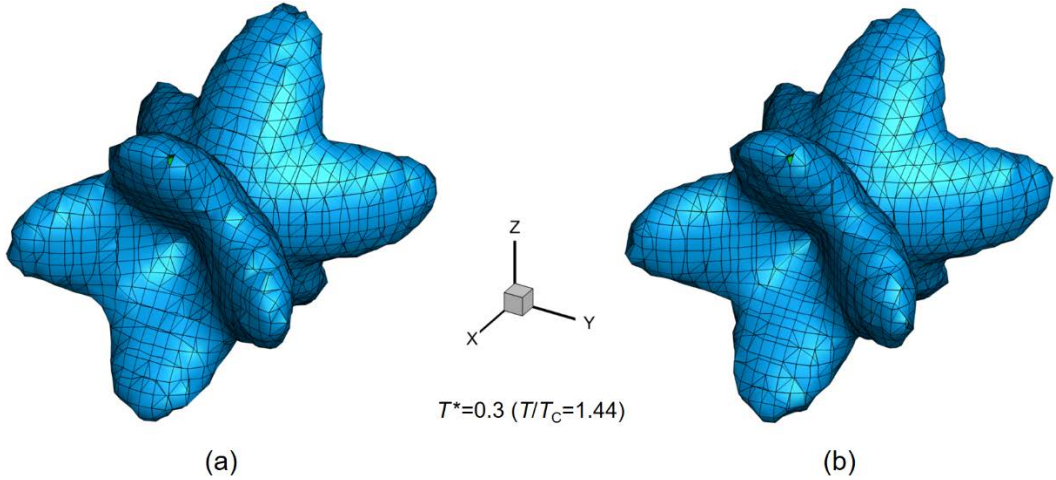

**Figure 6.** Simulated diffuse scattering around ($H$00) Bragg reflection peak at $T^*$=0.3 ($T$=1.44$T_\mathrm{C}$) computed by using (a) full formula $I_{\mathrm{diff}}(\mathbf{K};\mathbf{k}) \propto |(\mathbf{K}+\mathbf{k})\cdot\tilde{\mathbf{v}}(\mathbf{k})|^2$ in Eq. (17) and (b) simplified formula $I_{\mathrm{diff}}(\mathbf{K};\mathbf{k}) \propto |\mathbf{K}\cdot\tilde{\mathbf{v}}(\mathbf{k})|^2$ in Eq. (18).

Development of short-range order prior to phase transition producing long-range order is a phenomenon of fundamental importance. In particular, the displacement short-range ordering (i.e., spatial correlation in the heterogeneous lattice displacements) in the pre-martensitic austenite state is an interesting phenomenon. Existence of such displacement short-range ordering produces diffuse scattering in diffraction at intensities away from sharp Bragg reflection peaks, as described in Section 2.3. Figure 6(a) shows the simulated diffuse scattering around ($H$00) Bragg reflection peak at $T^*$=0.3 ($T$=1.44$T_\mathrm{C}$) computed by using Eq. (17), i.e., $I_{\mathrm{diff}}(\mathbf{K};\mathbf{k}) \propto |(\mathbf{K}+\mathbf{k})\cdot\tilde{\mathbf{v}}(\mathbf{k})|^2$. It is worth noting that, in order to directly link the simulated diffuse scattering to the experimentally measured diffuse scattering presented in Figures 1 and 2, a time averaging of the simulated diffuse scattering intensity over a sufficient number of Monte Carlo steps (1000 in this work) after reaching thermodynamic equilibrium is performed (Figure 6 indeed shows such time-averaged diffuse scattering), which is necessitated by both the relatively small computational supercell (32×32×32 used in this work) affordable to computer simulations and the relatively long exposure time required to experimentally measure the weak diffuse scattering intensity (discussed in Section 2.4). Figure 7 demonstrates the simulated instantaneous diffuse scattering (without time averaging) around ($H$00) Bragg reflection peak at $T^*$=0.3 ($T$=1.44$T_\mathrm{C}$) and the corresponding equilibrium microstructures at different simulation time $t^*$ (i.e., Monte Carlo step), after the system equilibrates during $t^*=0$ to 1000. As expected, the instantaneous diffuse scattering intensity shown in Figure 7 is very noisy due to the poor statistics associated with the small computational supercell affordable to the Monte Carlo simulations. It is observed that, more importantly, the time averaging procedure very effectively improves the statistics and the time-averaged diffuse scattering intensity shown in Figure 6 achieves significantly improved signal-to-noise ratio and exhibits the major features in agreement with the experimentally measured diffuse scattering shown in Figure 1. In this paper, simulation results of the time-averaged diffuse scattering are presented unless otherwise explicitly specified such as in Figure 7.

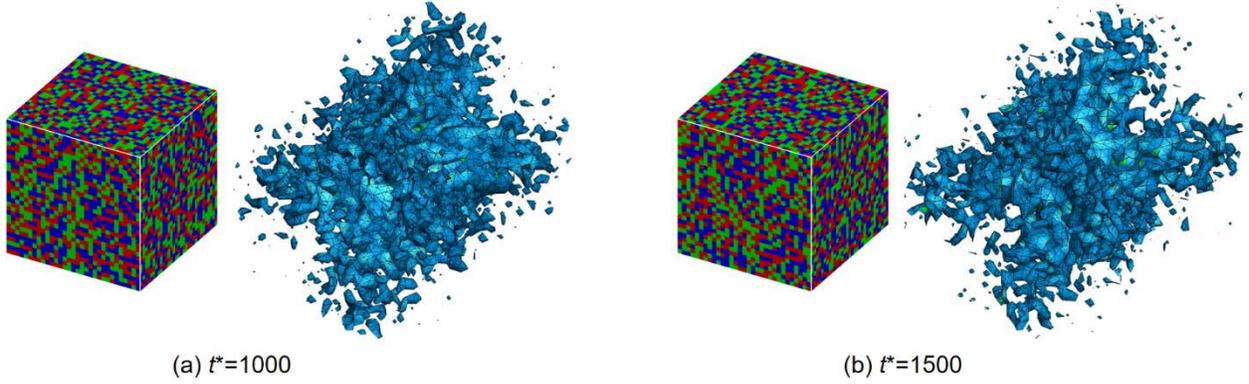

(a) $t^*=1000$    (b) $t^*=1500$

**Figure 7.** Simulated instantaneous diffuse scattering around ($H$00) Bragg reflection peak at $T^*=0.3$ ($T=1.44T_C$) corresponding to equilibrium microstructures at different simulation time (i.e., Monte Carlo step): (a) $t^*=1000$ and (b) $t^*=1500$.

It is also worth noting that the diffuse scattering is dominated by $\mathbf{K}$ and $\tilde{\mathbf{v}}(\mathbf{k})$, while the appearance of $\mathbf{k}$ in Eq. (17) accounts for the effect of local volume strain [22]. The volume strain effect is small in the case considered here and can be safely neglected, thus reducing Eq. (17) to the simplified formula:

$$I_{\text{diff}}(\mathbf{K};\mathbf{k}) \propto \left|\mathbf{K}\cdot\tilde{\mathbf{v}}(\mathbf{k})\right|^2 \qquad (18)$$

To compare with Figure 6(a), Figure 6(b) shows the simulated diffuse scattering around the same Bragg reflection peak at the same temperature computed by using Eq. (18), which demonstrates the accuracy and validity of the simplified formula. Eq. (18) makes it easier to reveal the physical implications of the diffuse scattering. As one example, consider the diffuse scattering around a ($H$00) Bragg reflection peak, where, according to Eq. (18), $I_{\text{diff}} \propto \left|\tilde{v}_x(\mathbf{k})\right|^2 = \tilde{v}_x(\mathbf{k})\tilde{v}_x(-\mathbf{k})$. It is readily shown that $I_{\text{diff}} \propto \tilde{P}(\mathbf{k}) = \int P(\boldsymbol{\rho})e^{-i\mathbf{k}\cdot\boldsymbol{\rho}}d^3\rho$, where

$$P(\boldsymbol{\rho}) = \int v_x(\mathbf{r})v_x(\mathbf{r}+\boldsymbol{\rho})d^3r \qquad (19)$$

is the auto-correlation function of the displacement field $v_x(\mathbf{r})$ in the lattice. This displacement field is generated by both the spontaneous lattice misfit strain of the martensite orientation variants and the elastic strain to maintain the lattice coherency. Figures 8(a) and (b) show the auto-correlation $P(\boldsymbol{\rho})$ obtained from the inverse Fourier transform of the instantaneous diffuse scattering intensity shown in Figures 7(a) and (b), respectively, which, as expected, are also noisy due to the poor statistics within small computational supercell. Figure 8(c) shows the time-averaged auto-correlation $\langle P(\boldsymbol{\rho})\rangle = \int \langle v_x(\mathbf{r})v_x(\mathbf{r}+\boldsymbol{\rho})\rangle d^3r$, which is obtained from the time-averaged diffuse scattering intensity shown in Figure 6. As discussed above, the time averaging effectively improves the statistics to achieve the results equivalent to that in adequately large systems (such as in experiments). The auto-correlation function $P(\boldsymbol{\rho})$ is analogous to the Patterson function of the electron density field in a crystal structure [25] and also the pair distribution function of the atoms in a material [26]. Unlike in long-range ordered systems where sharp peaks appear in the Patterson function and pair distribution function, which correspond to well-defined interatomic distance vectors in ordered atomic arrangements, the auto-correlation function $P(\boldsymbol{\rho})$ of the heterogeneous displacement field $v_x(\mathbf{r})$ in the disordered austenite state exhibits significant values only at small distance $\rho$ and vanishes rapidly with increasing distance, as shown in Figure 8, which characterizes short-range ordering in the displacement field. The displacement short-range ordering is dictated by the elastic interactions among the orientation variants: although the thermal fluctuations at temperature above the phase transition prevent a formation of long-range ordered configurations, the orientation variants develop a spatial correlation over short distance to reduce the elastic interaction energy, despite their apparently disordered distribution. Additionally, $P(\boldsymbol{\rho})$ exhibits a pronounced directional dependence, i.e., stronger correlation in $\langle 110\rangle$ directions, indicating displacement short-range ordering over a longer distance in $\langle 110\rangle$ directions. Such anisotropic features are also exhibited in the diffuse scattering shown in Figure 6, where stronger diffuse scattering intensity is produced along $\langle 110\rangle$ directions. It has been shown that such $\langle 110\rangle$ diffuse scattering rods are attributed to the displacement waves whose wavevectors are along $\langle 110\rangle$ directions and displacement vectors are along $\langle 1\bar{1}0\rangle$ directions [22]. Indeed, the Fourier transform $\tilde{\mathbf{v}}(\mathbf{k})$ given in Eqs. (14) and (15) is the plane wave representation of the heterogeneous displacement field $\mathbf{v}(\mathbf{r})$. It is worth noting that such displacement waves correspond to $\{110\}|\langle 1\bar{1}0\rangle$ shear strains and are associated with the twinning deformations on $\{110\}$ planes that transform one orientation variant into another variant of the tetragonal martensite.

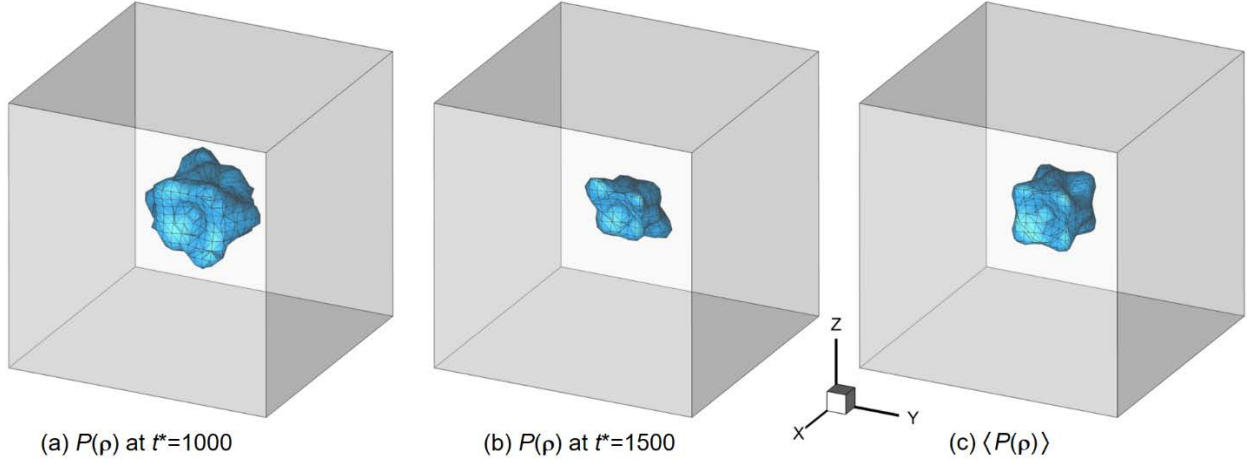

**Figure 8.** Simulated auto-correlation of displacement field $v_x(\mathbf{r})$ in equilibrium microstructures at $T^*=0.3$ ($T=1.44T_C$): (a) $t^*=1000$, (b) $t^*=1500$, and (c) time-averaged. Isosurface at 10% maximum cross-correlation value is visualized within 32×32×32 computational supercell to illustrate the nature of displacement short-range ordering.

### 3.4. Bragg Peak Dependence of Diffuse Scattering and Extinction Rule

According to Eq. (18) and analogous to Eq. (19), the diffuse scattering around a Bragg reflection peak with a general index (*HKL*) reflects the spatial correlation of the displacement field $\mathbf{v}(\mathbf{r})$ involving all three displacement components, i.e., $Hv_x(\mathbf{r})+Kv_y(\mathbf{r})+Lv_z(\mathbf{r})$, which indicates a strong dependence of the diffuse scattering on the Bragg peak index (*HKL*) or the reciprocal lattice vector **K** as described in Eqs. (17) and (18). Figure 9 shows the simulated diffuse scattering around different Bragg reflection peaks at temperature $T^*=0.25$ ($T=1.20T_C$), which demonstrates the (*HKL*) or **K** dependence of the diffuse scattering in agreement with the experimentally measured diffuse scattering shown in Figure 2. There are a total of 6 diffuse scattering rods (or 12 diffuse scattering spikes, i.e., every rod consists of two spikes pointing in opposite directions), each along one of $\langle 110 \rangle$ directions. It is worth noting that not all these diffuse scattering rods and spikes appear around every Bragg peak. As a matter of fact, individual diffuse scattering rods and spikes exhibit different relative intensities around different Bragg peaks and even completely disappear around certain Bragg peaks. Such extinction rule of diffuse scattering is described by Eq. (18). It has been shown that the heterogeneous displacement field $\mathbf{v}(\mathbf{r})$ producing such $\langle 110 \rangle$ diffuse scattering can be expressed as a sum of 12 branches of displacement plane waves with wavevectors along $\langle 110 \rangle$ directions and displacement vectors along $\langle 1\bar{1}0 \rangle$ directions [11,22]. In particular, the extinction rule of diffuse scattering allows us to associate each diffuse scattering rod to a particular displacement plane wave branch. For example, the diffuse scattering rod along $[110]$ direction appears around (100) Bragg peak but disappears around (110) and (001) Bragg peaks, as shown in Figure 9. The displacement plane wave branch producing the $[110]$ diffuse scattering rod has wavevectors along $[110]$ direction, i.e., $\tilde{\mathbf{v}}(\mathbf{k})$ is nonzero along $\mathbf{k}$ parallel to $[110]$ as required by Eq. (18), and displacement vectors along $[1\bar{1}0]$ direction, i.e., $\mathbf{v} = (v_x, -v_x, 0)$, where $\mathbf{K} \cdot \mathbf{v} = 0$ at $(HKL)=$(110) and (001) leading to extinction of diffuse scattering around (110) and (001) Bragg peaks. Around $(H00)$ Bragg peaks as shown in Figurers 6 and 9, the 4 diffuse scattering rods respectively along $[110]$, $[1\bar{1}0]$, $[101]$, and $[10\bar{1}]$ appear with equal intensity, while the 2 diffuse scattering rods respectively along $[011]$ and $[01\bar{1}]$ completely disappear, since the displacement vectors in the latter two branches are respectively along $[01\bar{1}]$ and $[011]$ directions, where $\mathbf{K} \cdot \mathbf{v} = 0$ for $(H00)$ Bragg peaks.

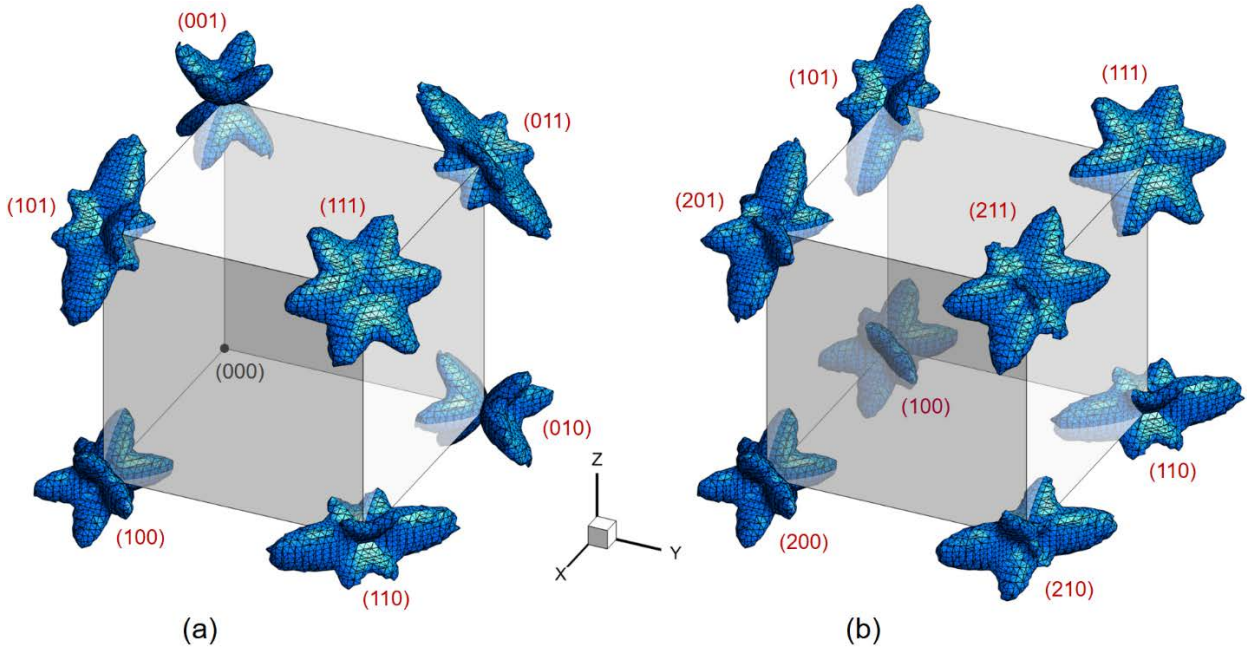

**Figure 9.** Simulated diffuse scattering around different Bragg reflection peaks (*HKL*) at temperature $T^*=0.25$ ($T=1.20T_C$).

## 3.5. Temperature Dependence of Diffuse Scattering and Displacement Short-Range Ordering

Figure 10 shows the simulated diffuse scattering around ($H$00) Bragg reflection peak at different temperatures above phase transition temperature $T_C^* = 0.209$, together with the representative equilibrium microstructures at the corresponding temperatures. The diffuse scattering becomes stronger upon cooling towards the phase transition temperature, in agreement with the experimentally measured diffuse scattering shown in Figure 1. As discussed in Section 3.3, the temperature dependence of the diffuse scattering manifests the temperature dependence of the displacement short-range ordering: with decreasing temperature, the correlation length of the displacement short-range ordering becomes longer, since the role of elastic interactions among the orientation variants characterized by $\Delta E_{el}$ becomes more prominent with respect to the thermal energy characterized by $k_B T$ at lower temperature. Despite the apparent resemblance in the disordered microstructures in the austenite state at different temperatures shown in the second row of Figure 10, the diffuse scattering shown in the first row of Figure 10 provides a direct means to quantitatively characterize the difference in the spatial correlation and short-range ordering of the heterogeneous displacement field at different temperatures in the pre-martensitic austenite state.

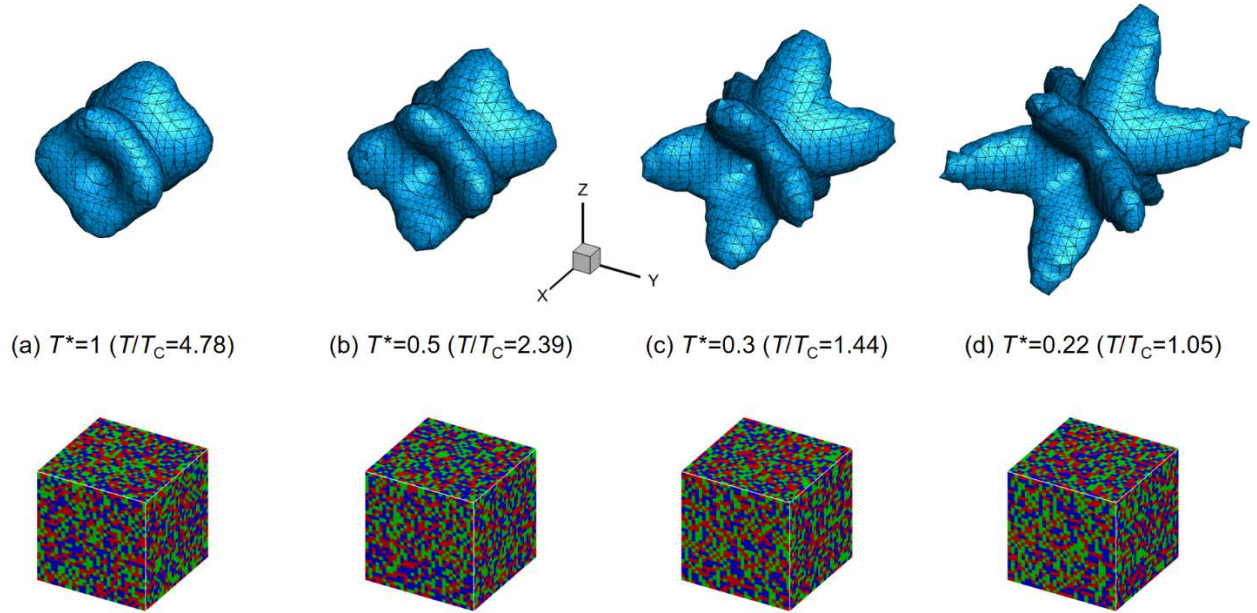

(a) $T^*=1$ ($T/T_C=4.78$)  (b) $T^*=0.5$ ($T/T_C=2.39$)  (c) $T^*=0.3$ ($T/T_C=1.44$)  (d) $T^*=0.22$ ($T/T_C=1.05$)

**Figure 10.** Simulated diffuse scattering around ($H$00) Bragg reflection peak at different temperatures above phase transition temperature $T_C^* = 0.209$: (a) $T^*=1$, (b) $T^*=0.5$, (c) $T^*=0.3$, and (d) $T^*=0.22$. Representative equilibrium microstructures at corresponding temperatures are shown in the second row.

### 3.6. Effects of Elastic Anisotropy and Softening of Shear Modulus $C'$

In this section, elastic anisotropy is considered. It is worth noting that while the shear modulus $C = C_{44}$ on (100) plane is relatively insensitive to temperature and remains approximately constant, the shear modulus $C' = (C_{11} - C_{12})/2$ on (110) plane usually exhibits significant softening upon cooling in pre-martensitic austenite state. Thus, the anisotropy factor defined as Zener ratio between the two shear moduli, i.e., $A = C/C'$ [27], reflects the degree of elastic softening in the shear modulus $C'$. Based on the results from a series of Monte Carlo simulations, Figure 11(a) shows the phase transition temperature $T_C^*$ as a function of the reciprocal anisotropy factor $A^{-1} = C'/C$, which is proportional to the shear modulus $C'$. It is observed that elastic softening in the shear modulus $C'$ leads to decrease in the phase transition temperature. This is caused by decrease in the strength of elastic interactions among the orientation variants: as shown in Eqs. (6) and (12), the elastic interaction energy is proportional to $C'$, and thus elastic softening in $C'$ leads to decreased value of $C'$ and weaker elastic interactions, resulting in more prominent role of thermal fluctuations and decrease in the phase transition temperature. Therefore, the elastic softening in the shear modulus $C'$ stabilizes the disordered state (austenite). On the other hand, elastic hardening in $C'$ leads to increase in the phase transition temperature. Despite the fact that the elastic interaction energy is proportional to $C'$, the phase transition temperature $T_C^*$ is not linearly proportional to $C'$ as shown in Figure 11(a), because the elastic interaction energy also depends on the anisotropy factor through $a = A - 1$ as shown in Eqs. (7)-(9).

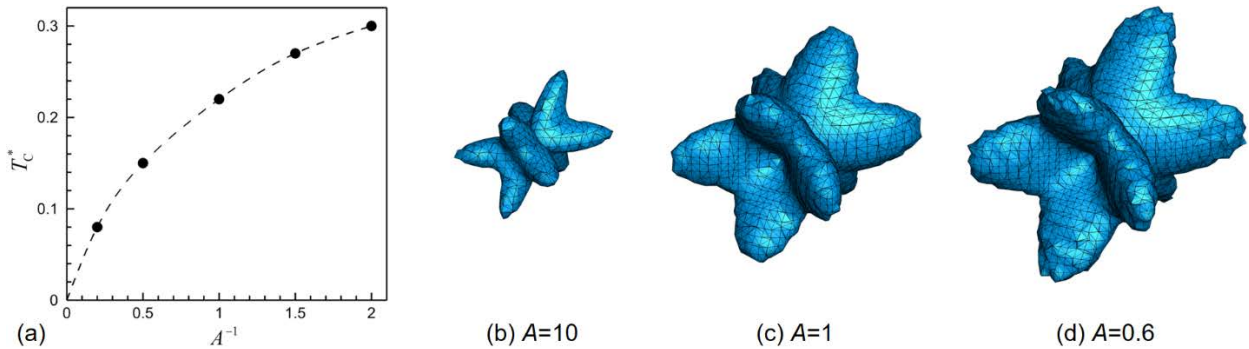

**Figure 11.** Effects of elastic anisotropy. (a) Simulated dependence of phase transition temperature on reciprocal anisotropy factor $A^{-1}$ (proportional to shear modulus $C'$), where dashed line is a guide to the eye. Simulated diffuse scattering around ($H$00) Bragg reflection peak at temperature $T^*$=0.3 with different elastic anisotropy factors: (b) $A$=10 corresponding to $C'$ softening; (c) $A$=1 corresponding to elastic isotropy; and (d) $A$=0.6 corresponding to $C'$ hardening.

Figures 11(b)-(d) show the simulated diffuse scattering around ($H$00) Bragg reflection peak at temperature $T^*$=0.3 with different elastic anisotropy factors (i.e., different degrees of elastic softening in the shear modulus $C'$). It is observed that, at the same temperature $T^*$=0.3, softened shear modulus $C'$ leads to weaker diffuse scattering. This is caused by the weaker elastic interactions among the orientation variants due to decreased $C'$ and thus weaker spatial correlation in the heterogeneous displacement field. Also, the same temperature $T^*$=0.3 corresponds to different "homologous" temperatures $T_H = T^*/T_C^*$ in the systems with different phase transition

temperatures $T_\text{C}^*$ due to different elastic anisotropy factors (different degrees of elastic softening in the shear modulus $C'$): according to $T_\text{C}^*$ shown in Figure 11(a), $T_\text{H} \approx 7$ in the softening case shown in Figure 11(b), and $T_\text{H} \approx 1.1$ in the hardening case shown in Figure 11(d), while $T_\text{H} = 1.44$ in the isotropy case shown in Figure 11(c). In particular, the hardening case of $A$=0.6 corresponds to the lowest "homologous" temperature $T_\text{H} \approx 1.1$ among the three cases and consequently exhibits the strongest diffuse scattering among the three cases. It is worth noting that the elastic softening in the shear modulus $C'$ leads to sharper diffuse scattering rods and spikes as shown in Figure 11 (b), in better agreement with the diffuse scattering feature shown in Figure 1 that is experimentally measured in Ni$_{49.90}$Mn$_{28.75}$Ga$_{21.35}$ single crystal that undergoes significant $C'$ softening in the pre-martensitic austenite state. Softening in $C'$ promotes $\{110\}|\langle 1\bar{1}0\rangle$ displacement plane waves that correspond to shear strains on $\{110\}$ planes, resulting in sharper diffuse scattering rods and spikes confined in $\langle 110 \rangle$ directions.

As discussed above, elastic softening tends to stabilize the high-temperature disordered austenite phase and decrease the martensitic transformation temperature. It is worth noting that such an effect can be explained in terms of increased entropy and thus decreased free energy in the disordered austenite phase. In the quasi-spin Ising model and Monte Carlo simulations presented above, the increased entropy effect is realized by more frequent switching of the orientation variants and shorter correlation length of the heterogeneous displacement field resulting from the weaker elastic interactions due to the softer shear modulus $C'$. This effect is analogous to the stabilization of the austenite phase by a more general form of elastic softening, namely, phonon softening, where the shear modulus softening corresponds to the long-wavelength limit of transverse acoustic phonon softening. It is well known that softened phonons stabilize the phases by increasing the lattice vibrational entropy [28]. The analogy between the two mechanisms, i.e., shear modulus softening and phonon softening, goes even further in the case of cubic→tetragonal martensitic transformations. In austenitic systems that undergo martensitic transformation into tetragonal phases, the transverse acoustic TA$_2$ phonon branches usually exhibit softening upon cooling towards the phase transition temperature. The TA$_2$ phonons are characterized by wavevectors along $\langle 110 \rangle$ directions and polarization (displacement) vectors along $\langle 1\bar{1}0 \rangle$ directions [11], which are the same as the displacement plane waves discussed in Sections 3.3 and 3.4. Indeed, the TA$_2$ phonons, when going to long-wavelength limit and undergoing complete softening, are believed to be responsible for the formation of tetragonal orientation variants during the martensitic transformation. Therefore, the frequent switching of the orientation variants in the Monte Carlo simulations and the associated short-range correlated evolution of the heterogeneous displacement field, which can be represented in the form of $\langle 110 \rangle|\langle 1\bar{1}0 \rangle$ displacement plane waves as discussed in Section 3.3, mimic the TA$_2$ phonons in the pre-martensitic austenite state,

where both stabilize the austenite phase, decrease the phase transition temperature, and condense to form tetragonal orientation variants during the martensitic transformation.

## 5. Summary

In this work, we focus on the intrinsic behaviors of a defect-free crystal that undergoes cubic-to-tetragonal martensitic transformation to gain insight into the diffuse scattering phenomenon in the pre-martensitic austenite state. To this end, a quasi-spin Ising model of ferroelastic phase transition is developed and employed to perform atomic-scale Monte Carlo simulation of thermoelastic martensitic transformation. The quasi-spin variable associated with the lattice sites characterizes the local unit cells of the orientation variants of the ground-state martensite phase, which interact with each other through long-range elastic interactions. The atomic-scale heterogeneous lattice displacements deviating from the average lattice sites are treated effectively in a manner of displacement plane waves that mimic acoustic phonons relevant to the martensitic transformation. The diffuse scattering is correlated to the short-range ordering of the atomic-scale heterogeneous lattice displacements that develop in the pre-martensitic austenite crystal lattice prior to the development of long-range order during martensitic transformation. In particular, the $\langle 110 \rangle$ diffuse scattering rods and spikes are attributed to the displacement plane waves with wavevectors along $\langle 110 \rangle$ directions and displacement vectors along $\langle 1\bar{1}0 \rangle$ directions, corresponding to $\{110\}|\langle 1\bar{1}0 \rangle$ shear strains associated with the twinning deformations on $\{110\}$ planes that transform one orientation variant into another variant of the tetragonal martensite. The effects of temperature, elastic anisotropy, and shear modulus softening on the diffuse scattering and displacement short-range ordering are investigated. The $\langle 110 \rangle|\langle 1\bar{1}0 \rangle$ displacement plane waves play a dominant role analogous to the transverse acoustic $TA_2$ phonons, both of which upon elastic softening stabilize the austenite phase, decrease the phase transition temperature, and condense to form tetragonal orientation variants during the martensitic transformation. The simulated diffuse scattering is compared and agrees with the complementary synchrotron X-ray single-crystal diffuse scattering experiment.

**Acknowledgments**

Support from NSF DMR-1506936 is acknowledged. Computer simulations were performed on XSEDE supercomputers. Use of Advanced Photon Source at Argonne National Laboratory was supported by DOE DE-AC02-06CH11357.

**References**

[1]   M.E. Fine, M. Meshii, C.M. Wayman, Martensitic Transformation (Academic Press, New York, 1978).

[2]  Z. Nishiyama, Martensitic Transformation (Academic Press, New York, 1978).

[3]  A.G. Khachaturyan, Theory of Structural Transformations in Solids (John Wiley & Sons, New York, 1983).

[4]  E.K.H. Salje, Phase Transitions in Ferroelastic and Co-elastic Crystals (Cambridge University Press, Cambridge, 1990).

[5]  N. Nakanishi, Elastic constants as they relate to lattice properties and martensite formation, Prog. Mater. Sci. 24 (1980) 143.

[6]  K. Otsuka, T. Kakeshita, Science and technology of shape-memory alloys: new developments, MRS Bulletin 27 (2002) 91.

[7]  G.R. Barsch, J.A. Krumhansl, L.E. Tanner, M. Wuttig, A new view on martensitic transformations, Scripta Metall. 21 (1987) 1257.

[8]  L.E. Tanner, W.A. Soffa, Pretransformation behavior related to displacive transformations in alloys, Metall. Trans. A 19 (1988) 158.

[9]  L.E. Tanner, M. Wuttig, Workshop on first-order displacive phase transformations: review and recommendations, Mater. Sci. Eng. A 127 (1990) 137.

[10] J.A. Krumhansl, Multiscale science: materials in the 21$^{st}$ century, Mater. Sci. Forum 327-328 (2000) 1.

[11] Y.M. Jin, Y.U. Wang, Y. Ren, Theory and experimental evidence of phonon domains and their roles in pre-martensitic phenomena, npj Comput. Mater. 1 (2015) 15002.

[12] Y.M. Jin, A. Artemev, A.G. Khachaturyan, Three-dimensional phase field model of low-symmetry martensitic transformation in polycrystal: simulation of $\zeta_2'$ martensite in AuCd alloys, Acta Mater. 49 (2001) 2309.

[13] Y.U. Wang, Y.M. Jin, A.G. Khachaturyan, The effects of free surfaces on martensite microstructures: 3d phase field microelasticity simulation study, Acta Mater. 52 (2004) 1039.

[14] P. Alippi, P.M. Marcus, M. Scheffler, Strained tetragonal states and Bain paths in metals, Phys. Rev. Lett. 78 (1997) 3892.

[15] F. Maresca, V.G. Kouznetsova, M.G.D. Geers, W.A. Curtin, Contribution of austenite-martensite transformation to deformability of advanced high strength steels: from atomistic mechanisms to microstructural response, Acta Mater. 156 (2018) 463.

[16] S. Kartha, J.A. Krumhansl, J.P. Sethna, L.K. Wickham, Disorder-driven pretransitional tweed pattern in martensitic transformations, Phys. Rev. B 52 (1995) 803.

[17] M. Blume, V.J. Emery, R.B. Griffiths, Ising model for the λ transition and phase separation in $He^3$-$He^4$ mixtures, Phys. Rev. A 4 (1971) 1071.

[18] E.D. Belokolos, A.Yu. Gaevsky, The axial Ising model for martensitic transformations and pseudoelasticity, J. Phys. Chem. Solids 50 (1989) 1199.

[19] S.P. Shrestha, D. Pandey, Application of 1d kinetic Ising model for studying the evolution of diffuse scattering during HCP (2H) to FCC (3C) martensitic transformation, Europhysics Lett. 34 (1996) 269.

[20] G.P.P. Pun, V. Yamakov, Y. Mishin, Interatomic potential for the ternary Ni-Al-Co system and application to atomistic modeling of the B2-L1$_0$ martensitic transformation, Modelling Simul. Mater. Sci. Eng. 23 (2015) 65006.

[21] N. Metropolis, A.W. Rosenbluth, M.N. Rosenbluth, A.H. Teller, E. Teller, Equation of state calculations by fast computing machines, J. Chem. Phys. 21 (1953) 1087.

[22] Y.M. Jin, Y.U. Wang, Diffuse scattering intensity distribution associated with static and dynamic atomic position fluctuations, JOM 64 (2012) 161.

[23] T.L. Cheng, F.D. Ma, J.E. Zhou, G. Jennings, Y. Ren, Y.M. Jin, Y.U. Wang, In-situ three-dimensional reciprocal space mapping of diffuse scattering intensity distribution and data analysis for precursor phenomenon in shape memory alloy, JOM 64 (2012) 167.

[24] L.D. Landau, E.M. Lifshitz, Statistical Physics (Pergamon Press, Oxford, 1980).

[25] A.L. Patterson, A Fourier series method for the determination of the components of interatomic distances in crystals, Phys. Rev. 46 (1934) 372.

[26] B.E. Warren, X-Ray Diffraction (Addison-Wesley Publishing: Reading, Massachusetts, 1969).

[27] C. Zener, Contributions to the theory of beta-phase alloys, Phys. Rev. 71 (1947) 846.

[28] M. Born, K. Huang, Dynamical Theory of Crystal Lattices (Oxford University Press, Oxford, 1954).